\documentclass[preprint,sort&compress]{elsarticle}
\usepackage{lineno,hyperref}
\usepackage{epstopdf, epsfig}
\usepackage[export]{adjustbox}
\usepackage[amssymb]{SIunits}
\usepackage{xcolor} 
\usepackage{amsfonts}
\usepackage{latexsym}
\usepackage{amssymb}
\usepackage[tight,footnotesize]{subfigure}
\usepackage{booktabs}
\usepackage{multirow}
\usepackage[utf8]{inputenc}
\usepackage{longtable}
\usepackage{array}
\usepackage{wasysym}
\usepackage{tikz}
\usepackage{wasysym}
\usepackage{graphicx}
\usepackage[section]{placeins}
\usepackage{natbib}

\usepackage{microtype}

\definecolor{height_max}{rgb}{0.6350, 0.0780, 0.1840}
\definecolor{height_mid}{rgb}{0.4660, 0.6740, 0.1880}
\definecolor{height_min}{rgb}{0.0, 0.45, 0.73}

\definecolor{deepcarrotorange}{rgb}{0.91, 0.41, 0.17}
\definecolor{mtlb_orange}{rgb}{0.9290, 0.6940, 0.1250}
\definecolor{mtlb_blue}{rgb}{0.0, 0.45, 0.73}
\definecolor{mtlb_green}{rgb}{0.4660, 0.6740, 0.1880}
\definecolor{mtlb_red}{rgb}{0.6350, 0.0780, 0.1840}
\definecolor{mtlb_purple}{rgb}{ 0.4940, 0.1840, 0.5560}

\modulolinenumbers[0]
\usepackage{lineno}

\journal{Chemical Engineering Science}

\begin{document}

\begin{frontmatter}

\title{Experimental investigation of heat transport in inhomogeneous bubbly flow}

\author[pof]{Biljana Gvozdi\'{c}}
\author[pof]{On--Yu Dung}
\author[elise,pof]{Elise Alm\'{e}ras}
\author[pof]{Dennis P. M. van Gils}
\author[pof]{Detlef Lohse}
\author[pof]{Sander G. Huisman\corref{cor1}}
\ead{s.g.huisman@gmail.com}
\author[chao,pof]{Chao Sun\corref{cor1}}
\ead{chaosun@tsinghua.edu.cn}

\address[pof]{Physics of Fluids Group, J. M. Burgers Center for Fluid Dynamics and Max Planck Center Twente, Faculty of Science and Technology, University of Twente, P.O. Box 217, 7500 AE Enschede, The Netherlands}
\address[elise]{Laboratoire de G\'{e}nie Chimique, UMR 5503, CNRS-INP-UPS, 31106 Toulouse, France}
\address[chao]{Center for Combustion Energy, Key Laboratory for Thermal Science and Power Engineering of Ministry of Education, Department of Energy and Power Engineering, Tsinghua University, 100084 Beijing, China}

 \cortext[cor1]{Corresponding authors}

\begin{abstract}

In this work we study the heat transport in inhomogeneous bubbly flow.
The experiments were performed in a rectangular bubble column heated from one side wall and cooled from the other, with millimetric bubbles introduced through one half of the injection section (close to the hot wall or close to the cold wall).
We characterise the global heat transport while varying two parameters: the gas volume fraction $\alpha = 0.4\% -  5.1 \%$, and the Rayleigh number $Ra_H=4\times10^9-2.2\times10^{10}$. 
As captured by imaging and characterised using Laser Doppler Anemometry (LDA), different flow regimes occur with increasing gas flow rates. 
In the generated inhomogeneous bubbly flow there are three main contributions to the mixing: (\textit{i}) transport by the buoyancy driven recirculation, (\textit{ii}) bubble induced turbulence (BIT) and (\textit{iii}) shear-induced turbulence (SIT).  
The strength of these contributions and their interplay depends on the gas volume fraction which is reflected in the measured heat transport enhancement. 
We compare our results with the findings for heat transport in homogeneous bubbly flow from Gvozdi{\'c} \emph{et al.} (2018)\citep{gvozdic2018experimental}. 
We find that for the lower gas volume fractions ($\alpha<4\%$), inhomogeneous bubbly injection results in better heat transport due to induced large-scale circulation.
In contrast, for $\alpha>4\%$, when the contribution of SIT becomes stronger, but so does the competition between all three contributions, the homogeneous injection is more efficient.

\end{abstract}


\end{frontmatter}

\section{Introduction}

Injection of bubbles in a continuous liquid phase is widely used to enhance mixing without any additional mechanical parts. 
As a result, bubbly flows enhance heat and mass transfer and can therefore be found in various industrial processes such as synthesis of fuels and basic chemicals, emulsification, coating, fermentation, etc. 
In particular, to understand the effect of bubbles on heat transport, a variety of flow configurations have been used in previous works. 
These studies can be broadly classified based on (\textit{i}) the nature of forcing of the liquid, i.e. natural convection (liquid is purely driven by buoyancy) \texorpdfstring{\citep{kitagawa2008heat,kitagawa2009effects}} or forced convection (liquid is driven by both buoyancy and an imposed pressure gradient or shear) \citep{sekoguch1980forced, sato1981momentum, sato1981momentum2,dabiri2015heat}; and (\textit{ii}) based on the size of the injected bubbles, i.e. sub-millimetric bubbles \citep{kitagawa2008heat,kitagawa2009effects,kitagawa2013natural} to millimetric bubbles \citep{tokuhiro1994natural,deen2013direct}. 

Owing to the high complexity of the physical mechanism behind the bubble induced heat transfer enhancement, a systematic approach has to be taken when studying this phenomenon. 
Starting from a relatively simple case: bubbly flow in water combined with natural convection, Kitagawa \emph{et al.} (2013) \cite{kitagawa2013natural} studied the effect of bubble size on the heat transfer. 
They found that micro-bubbles (mean bubble diameter $d_{bub} = \unit{0.04}{\milli \meter}$) which form large bubble swarms close to the wall with significant wall normal motion, induce higher heat transfer enhancement as compared to sub-millimeter-bubbles ($d_{bub} = \unit{0.5}{\milli \meter}$), which have weak wakes and low bubble number density.

In our previous work \citep{gvozdic2018experimental} we studied heat transfer combined with natural convection with injection of millimetric bubbles  in water which due to their strong wake enhance the heat transport even more.
Those experiments were performed in a rectangular bubble column heated from one side and cooled from the other in order to understand the influence of homogeneously injected millimetric bubbles on the overall heat transport. 
The primary advantage with such a setup is that the dynamics of homogeneous bubbly flows has been adequately characterised and studied in the past 
\citep{risso2002velocity,riboux2010experimental,roghair2011energy,mercado2010bubble} and the flow without bubbles resembles the classical vertical natural convection system \citep{elder1965turbulent,markatos1984laminar,kimura1984boundary,belmonte1994temperature,ng2015vertical,shishkina2016thermal,ng2017changes}. 
The strength of the thermal driving of the fluid in such a system is characterised by the Rayleigh number which is the dimensionless temperature difference:
\begin{equation}
  Ra_H =  \frac{g  \beta  (\overline{T_h}-\overline{T_c})  H^3}{\nu   \kappa};
\end{equation}
and the dimensionless heat transfer rate, the Nusselt number:
\begin{equation}
 \overline{Nu} = \frac{Q / A}{\chi \ (\overline{T_h} - \overline{T_c})/ L  },
 \end{equation}
where $Q$ is the measured power supplied to the heaters, $\overline{T_h}$ and $\overline{T_c}$ are the mean temperatures (over space and time) of the hot wall and cold wall, respectively, $L$ is the length of the setup, $A$ is the surface area of the sidewall, $\beta$ is the thermal expansion coefficient, $g$ the gravitational acceleration, $\kappa$ the thermal diffusivity, and $\chi$ the thermal conductivity of water. 
Gvozdi{\'c} \emph{et al.}  \citep{gvozdic2018experimental}  found that homogeneous injection of bubbles in vertical natural convection can lead to a 20 times enhancement of the heat transfer compared to the corresponding flow with no bubbles. 
It was found that for $Ra_H=4.0 \times 10^9 - 2.2 \times 10^{10}$ and a gas volume fraction of $\alpha=0.5\% - 5\%$ the Nusselt number remained nearly constant for increasing $Ra_H$. 
Furthermore, Gvozdi{\'c} \emph{et al.} (2018) \citep{gvozdic2018experimental} found good agreement for the scaling of an effective diffusivity $D$ with the gas volume fraction $\alpha$ with the results of mixing of a passive scalar in a homogeneous bubbly flow \citep{almeras2015mixing}, i.e. roughly $D \propto \alpha^{1/2}$, which implies that the bubble-induced mixing is controlling the heat transfer.

With a goal to further enhance bubble induced heat transport in vertical natural convection, in this study we explore the influence of inhomogeneous bubble injection on the overall heat transfer. 
Previous studies have shown that inhomogeneous gas injection induces mean liquid circulations (large-scale coherent rolls) in bubbles columns \cite{almeras2018mixing, almeras2016scalar}. 
It is also known from classical Rayleigh-B{\'e}nard convection that aiding formation of the coherent structures can enhance heat transfer \citep{ahlers2009heat,lohse2010small}. 
In this work, we take advantage of both these phenomena and use the large-scale circulation generated by inhomogeneous bubble injection in a vertical natural convection setup to further enhance heat transport as compared to the case of homogeneous injection of bubbles. 
We use the same experimental setup as in Gvozdi{\'c} \emph{et al.} (2018) \citep{gvozdic2018experimental}, while we inject the bubbles through one half of the injection section, either close to the hot wall or close to the cold wall (see Figure \ref{fig1}).
We characterise the global heat transfer while varying  two parameters: the gas volume fraction $\alpha = 0.4\% -  5.1 \%$, and the Rayleigh number $Ra_H = 4 \times 10^9 - 2.2\times 10^{10}$.
We compare findings on global heat transfer for the cases of homogeneous bubble injection, injection close to the hot wall, and injection close to the cold wall. 
We further demonstrate the difference in the dynamics between lower gas volume fraction case  ($\alpha=0.4\%$) and higher gas volume fraction case ($\alpha=3.9\%$) by performing velocity profile measurements along the length of the setup at mid-height.

The paper is organised as follows. In section 2, we discuss the experimental set-up and the different flow configurations studied. Results on the liquid flow characterisation and on the global heat transfer enhancement are detailed in section 3 while concluding remarks are given in section 4.

\section{Experimental setup and instrumentation} \label{experiments}

\subsection{Experimental setup} 
\begin{figure}
\centering
\includegraphics[scale=0.85]{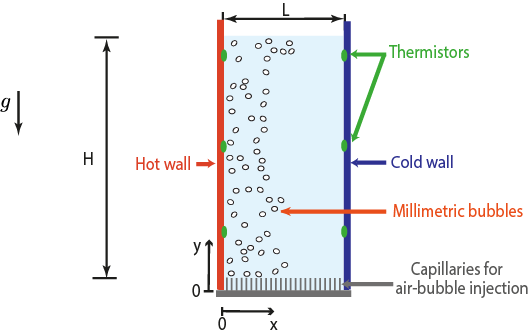}
\caption{Rectangular bubbly column heated from one sidewall and cooled from the other ($H = \unit{600}{\milli \meter}$, $L = \unit{230}{\milli \meter}$). Bubbles are injected either close to the hot wall or close to the cold wall, through 90 capillaries (out of 180 in total) placed at the bottom of the setup (inner diameter $\unit{0.21}{\milli \meter}$).}
\label{fig1}
\end{figure}

In figure \ref{fig1}, we show a schematic of the experimental setup. 
The apparatus consists of a rectangular bubble column, where the two main sidewalls of the setup ($ 600 \times \unit{230}{\milli \meter ^2}$) are made of $\unit{1}{\centi \meter}$ thick glass and the two (heated resp. cooled) sidewalls ($ 600 \times  \unit{60}{\milli \meter ^2}$) of $\unit{1.3}{\centi \meter}$ thick brass.
Heating is provided via Joule heaters placed on the brass sidewall, while cooling of the opposite brass wall is performed using a circulating water bath.
The temperature of these walls is monitored using thermistors. 
Millimetric bubbles are injected through 90 out of the total 180 capillaries at the bottom of the setup, either close to the hot wall or close to the cold wall.

\begin{figure*} 
\centering
\includegraphics[scale=0.7]{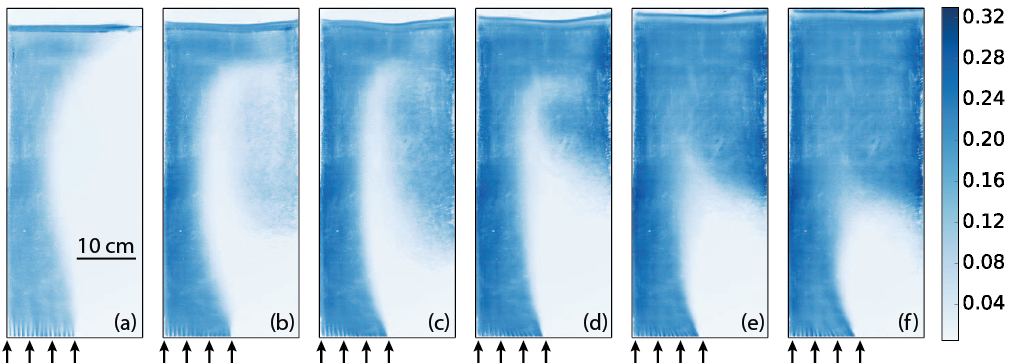}
\caption{Visualisation of preferential concentration of bubbles for each gas volume fraction: (a) $\alpha =0.4\%$, (b) $\alpha =0.9\%$, (c) $\alpha =1.4\%$, (d) $\alpha =2.3\%$, (e) $\alpha =3.9\%$, (f) $\alpha = 5.1\%$; arrows mark the gas injection. Colour corresponds to $\sigma/\langle\sigma_{bg}\rangle$, that is the  standard deviation of each pixel intensity $\sigma$ normalised by $\langle\sigma_{bg}\rangle$ the mean of the standard deviation of pixel intensity for a background image taken without the bubble injection.}

\label{regimes}
\end{figure*}

The global gas volume fraction is modulated between $0.4 \%$ to $5.1\%$ by varying the inlet gas flow rate.
Global gas volume fraction was estimated as an average elevation of the liquid at the top of the setup, which is measured by processing images captured using a Nikon D850 camera. 
We find that different flow regimes develop with increasing inlet gas flow rate. 
Movies capturing these regimes can be found in the Supplementary material.
In order to visualise the preferential concentration of the bubbles in figure \ref{regimes} we show the normalised standard deviation of each pixel in the movie frame converted to grayscale over around 1500 frames.
For low global gas volume fractions (around $0.4\%$) we visually observe that the bubbles rise without migrating to the opposite side (see Figure \ref{regimes}(a)), while the bubble stream bends due to liquid recirculation caused by the pressure gradient between the two halves of the setup, as it was previously observed by \cite{roig1998experimental}.
For a global gas volume fraction of approximately $1\%$, the bubbles start migrating to the opposite side (see Figure \ref{regimes}(b)), inducing a weak bubble circulation loop on the opposite half of the setup. 
This recirculation loop does not interfere with the main bubble stream. 
At an even higher gas volume fraction of around $2.3 \%$, a significant part of the bubble stream passes to the other half of the setup, and strongly interacts with the main bubble stream (see Figure \ref{regimes}(d)). 
The migrating bubbles form an unstable loop, which gets partially trapped by the main bubble stream and carried to the top of the setup. 
With increasing gas flow rate, the amount of bubbles passing from the injection side to the opposite half of the setup increases, as does the instability of the main bubble stream.
These regimes have significant influence on the heat transfer, which will be addressed later in section \ref{global}.

\subsection{Instrumentation for the gas phase characterisation} \label{gas_char}

\begin{figure} 
\centering
\includegraphics[scale=1]{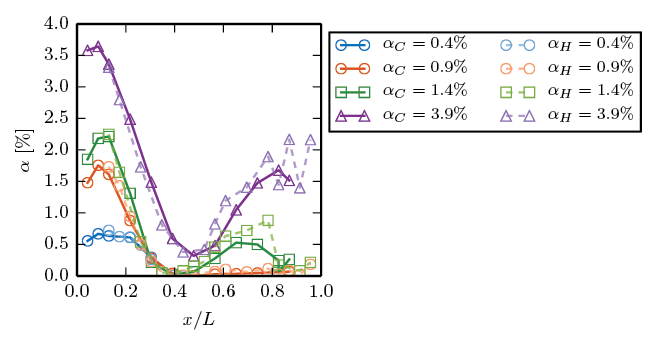}
\caption{Gas volume fraction profiles at the mid-height for injection close to the hot wall $\alpha_H$ and injection close to the cold wall $\alpha_C$ (cold wall profiles are mirrored around $x/L=0.5$ for the comparison).}
\label{gvf}
\end{figure}

In order to characterise the gas phase, we first perform a scan of the local gas volume fraction using a single optical fibre probe (for working principle see \cite{cartellier1990optical}). 
We perform measurements only at half-width of the column because the capilaries for bubble injection are arranged in 6 rows and 30 columns, and the width of the column is only $\unit{6}{\centi \meter}$.
In figure \ref{gvf}, we plot the local gas volume fraction measured at half-height $y/H = 0.5$, versus the normalised length for both hot-wall and cold-wall injection. 
We find that the profiles of the gas volume fraction do not differ significantly for the cases where the bubbles are injected close to the hot wall or close to the cold wall. 

\begin{figure} 
\centering
\includegraphics[scale=0.91]{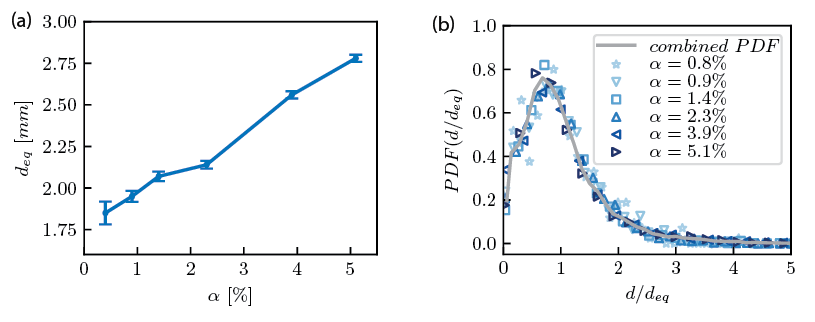}
\caption{(a) Mean bubble diameter $d_{eq}$ for different gas volume fractions $\alpha$, the error-bar presents the standard error of the mean bubble diameter. (b) Probability density function (PDF) of the normalised bubble diameter. The symbols present the PDF for each gas volume fraction, while the line presents a PDF calculated using data points for all studied gas volume fractions.}
\label{d_eq}
\end{figure}

The bubble diameter and bubble velocity were measured at half height using an in-house dual optical probe, which consists of two optical fibres placed one above the another. 
The bubble velocity is given as $V_{bub} = \delta / \Delta t$, where $\delta = \unit{4}{\milli \meter}$ is the vertical distance between the fibre tips and $\Delta t$ is the time interval between which one bubble successively pierces each fibre.
The diameter is estimated from the time during which the leading probe is in the gas phase. 
To ensure precise measurements of the bubble diameter and velocity, we perform measurements only in the injection stream. 
On the opposite side of the injection stream, bubbles move downwards which makes their accurate detection with a downward facing probe impossible. 
Bubble diameter measurements indicate an expected increasing trend with increasing gas volume fraction (see Figure \ref{d_eq} (a)), while the distribution of the normalised bubble diameter remains nearly the same for all the gas volume fractions with a standard deviation of 0.7 (see Figure \ref{d_eq} (b)).
The bubble velocity is in the range $V_{bub} = \unit{0.5 \pm 0.02}{\meter \per \second} $ for the given gas volume fraction span.
The bubble rising velocity in the injection leg of the setup is nearly constant as it can be expressed as $V_{bub}  = U + V_r$, and the mean rising liquid velocity $U$ increases (see Section \ref{velocity}) while the relative velocity $V_r$ decreases with increasing gas volume fraction \citep{riboux2010experimental}, thus compensating each other.

\subsection{Instrumentation for the liquid phase velocity measurements}
The vertical and the horizontal component of the liquid phase velocity are measured by means of Laser Doppler Anemometry (LDA) in backscatter mode. 
The flow is seeded with polyamid seeding particles (diameter $\unit{5}{ \micro \meter}$, density $1050\ kg/m^3 $). 
The LDA system used consists of DopplerPower DPSS (Diode Pumped Solid State) laser and a  Dantec  burst spectrum analyser (BSA).
It has been shown previously that the LDA in backscatter mode measures predominantly the liquid velocity \citep{mudde1997liquid,groen1999wobbling,vial2001influence}. 
Therefore, no post-processing was performed on the data. 
Measurements of 30 minutes were performed for each measurement point with a data rate of $\mathcal{O} (100)$ Hz. %

\subsection{Instrumentation for the heat flux measurements}

In order to characterise the global heat transport, namely to obtain $\overline{Nu}$ and  $Ra_H$, we measured the hot and cold wall temperatures (the control parameters), and the heat input to the system (the response parameters). 
Accordingly, resistances of the thermistors placed on the hot and cold walls and the heat power input were read out every 4.2 seconds using a digital multimeter (Keysight 34970A).
Operating temperatures for each Rayleigh number are given in table \ref{table:dT}.
The heat losses were estimated to be in the range of $3\%$ to $7\%$ by calculating convective heat transport rate from all outer surfaces of the setup if they are at $\unit{25}{^\circ \Celsius}$ and by measuring the power needed to maintain the temperature of the bulk constant ($T_{bulk} = \unit{25}{^\circ \Celsius}$) over 4 hours.
More details on the temperature control of the setup and measurements of the global heat flux are provided in \cite{gvozdic2018experimental}. 
The experimental data was acquired after a steady state was achieved in which the mean wall temperatures fluctuated less that $\pm 0.5$ K.
Time averaging of the instantaneous supplied heating power was then performed over a period of 3 hours.

\begin{table}
	\begin{center}
		\setlength\extrarowheight{6pt}

		\begin{tabular}{   r  || r|  r  | r | r| r | r| r |r|r }
			{$Ra_H \times 10^{-9}$}   &		{\ 4.0\ }	&{\ 5.3\ }		&	{\ 6.8\ }	&	{\ 9.1\ }	&	{\ 12.3\ } &	{\ 16.5\ }	&	{\ 22.4\ }	&	{\ 30.2\ }	&	\\ \hline
			{$\Delta T \ [K]$ }  &						{\ 2\ }		&	{\ 2.6\ }	&	{\ 3.3\ }	&	{\ 4.3\ }	&	{\ 5.6\ }	&	{\ 7.2\ } 		&	{\ 9.3\ } 	&	{\ 11.3\ }\\
		\end{tabular}
	\end{center}

	\caption{Operating values of $\Delta T$ at different $Ra_H$}
	\label{table:dT}
\end{table}

\section{Results} \label{results}

\subsection{Global heat transfer enhancement} \label{global}

\begin{figure*} 
\centering
\includegraphics[scale=1]{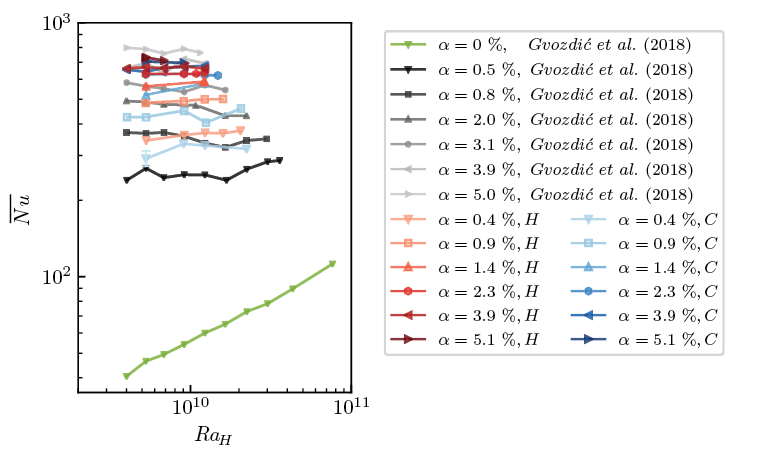}
\caption{Dependence of the Nusselt number $\overline{Nu}$ on the Rayleigh number $Ra_H$ for different gas volume fractions. Green data presents the single-phase case from \citep{gvozdic2018experimental}, black data the homogeneously injected bubbles from \citep{gvozdic2018experimental}, red data the injection of bubbles close to the hot wall (H) and blue data injection close to the cold wall (C).}
\label{Ra_Nu}
\end{figure*} 

We now analyse the heat transport in the presence of an inhomogeneous bubble swarm for gas volume fraction $\alpha$ ranging from $0.4\%$ to $5.1\%$ and the Rayleigh number ranging from $4 \times 10^9$ to $2.2\times 10^{10}$. 
Experiments performed in a previous study showed that in the single-phase case Nusselt increases with $Ra_H$  as $\overline{Nu} \propto {Ra_H}^{0.33}$, while in the case of homogeneous bubble injection $\overline{Nu}$ remains nearly constant with increasing $Ra_H$ \citep{gvozdic2018experimental}. 
In figure \ref{Ra_Nu}, along with the results previously obtained for the cases of single-phase flow and homogeneous bubble injection, we plot the Nusselt number versus $Ra_H$ for inhomogeneous injection. 
Similarly to the results for the homogeneous bubble injection, we find that  $\overline{Nu}$ remains independent of $Ra_H$ and is an order of magnitude higher  when compared to the single-phase case even if the bubbles are injected only through one half of the bubble injection section.

\begin{figure*} 
\centering
\includegraphics[scale=1]{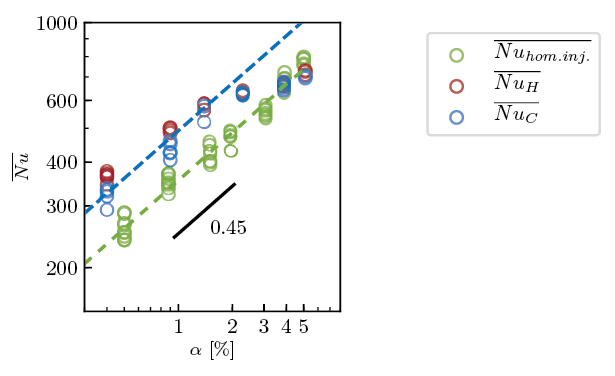}
\caption{Nusselt number $\overline{Nu}$ as a function of the gas volume fraction $\alpha$, subscripts: $C$ - injection close to the cold wall; $H$ - injection close to the hot wall, $hom.\ inj.$ -  homogeneous bubble injection. Hollow symbols show all the experimental measurements, and lines present the $\overline{Nu} \propto \alpha^{0.45}$ scaling. The crossover from strong enhancement by the inhomogeneous bubble injection for low $\alpha$ towards no enhancement for high $\alpha$ is clearly seen.}
\label{Nu_alpha}
 \end{figure*} 

Although Nusselt is not a function of Rayleigh in both cases, the heat transport enhancement mechanisms in case of inhomogeneous injection is different than the one present in case of homogeneously injected bubbles where $\overline{Nu} \propto \alpha^{0.45 \pm 0.025}$.
While in the case of a homogeneous bubble injection the mixing mechanism limiting the heat transport is bubble induced turbulence (BIT) \citep{gvozdic2018experimental}, the large-scale circulation of the liquid phase induced by inhomogeneous bubble injection leads to the occurrence of a shear layer between the fluid region injected with bubbles and its opposite side. 
Therefore, in the case of inhomogeneous bubble injection there are additional contributions to the mixing: the shear-induced turbulence (SIT) and mixing by large-scale liquid circulation. 
Given the difference in the fundamental mixing mechanisms for homogeneous injection and inhomogeneous injections, one can expect a difference in the scaling of the Nusselt number with $\alpha$.
Indeed, for gas volume fractions $\alpha \geq 1.4\%$  $\overline{Nu}$ does not follow the same trend with gas volume fraction as in the case of homogeneous injection, it seems to be less affected by changing $\alpha$ (see Figure \ref{Nu_alpha}).
However, for low gas volume fractions the scaling exponent of  $\overline{Nu}$ with $\alpha$ for inhomogeneous injection agrees well with homogeneous injection, though with an increased prefactor.
If we calculate the exact scaling exponent for inhomogeneous injection for gas volume fraction $\alpha \leq 1.4\%$, for hot wall injection we get $0.37 \pm 0.03$, for cold wall it is $0.41 \pm 0.08$. 
These exponents are comparable to the one for homogeneous injection.
It should be however taken into account that better precision of the exact scaling exponent would be obtained if the number of data points at low gas volume fractions would be greater, since in the present work only 3 gas volume fractions are available. 
The small observed difference in the exact scaling exponent between hot wall injection and cold wall injection will be addressed later.

\begin{figure} 
\centering
\includegraphics[scale=1]{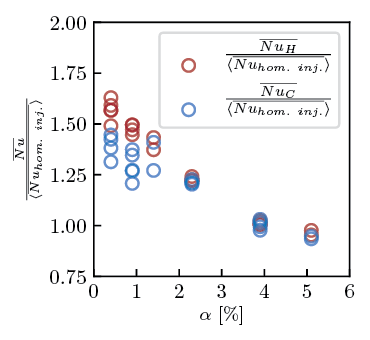}
\caption{The ratio of Nusselt for the inhomogeneous bubble injection and Nusselt for homogeneous injection as a function of the gas volume fraction.}
\label{Nu_alpha_HIdiv}
 \end{figure}

In order to understand the observed trend of Nusselt with increasing $\alpha$, we plot the ratio of Nusselt number for inhomogeneous injection to the corresponding Nusselt for the homogeneous injection for the same gas volume fraction in figure \ref{Nu_alpha_HIdiv}. 
The results indicate that the heat transfer enhancement (as compared to the homogeneous bubble injection) decreases with increasing gas volume fraction. 
Studies focused on characterisation of the mixing in similar inhomogeneous bubbly flow have shown that (\textit{i}) in case of  $\alpha < 3.5\%$ the mixing is enhanced using inhomogeneous injection \cite{almeras2016scalar} \cite{almeras2018mixing}, (\textit{ii}) for $\alpha < 3\%$ in the case of inhomogeneous bubbly flow where the buoyancy driven flow generates shear-induced turbulence involving wide range of scales from the from the size of the column to bubble diameter, the mixing time evolves as $\alpha^{-0.5}$ \cite{almeras2016scalar}.
Our results for low gas volume fraction agree well with these findings. 
On the other hand, at higher gas volume fractions the observed trend is changed.
Possible cause of decreased heat transport enhancement  at higher gas volume fractions is that the interacting mixing mechanisms might deteriorate one another, resulting in inhomogeneous bubble injection to be less effective that the homogeneous one for $\alpha>4\%$.

We now compare the global heat transfer for injection close to the hot wall and injection close to the cold wall. 
From figures \ref{Ra_Nu} and \ref{Nu_alpha} we find that there is almost no distinction between hot wall injection and cold wall injection for the same gas volume fraction for higher $\alpha$. 
The difference is more prominent for low gas volume fractions as compared to high gas volume fractions; namely, up to a critical value of the gas volume fraction $\alpha<1.4\%$ the Nusselt number for hot wall injection is slightly higher than the one for cold wall injection. 
One possible reason for the observed difference in the Nusselt number could be the different interaction between the rising bubbles and the falling cold boundary layer or rising hot boundary layer, namely the co-current flow of bubbles with the thermal boundary layer might perturb the boundary layer more than the counter-current flow.
However, we suspect that this is not the actual cause of the difference between hot wall injection and cold wall injection for two reasons: (\textit{i}) We do not see the difference between the different injection sides for higher gas volume fractions and  (\textit{ii}) the Nusselt number does not depend on the Rayleigh number, meaning that it is very likely that the perturbation of the boundary layers by the bubbles rising next to the wall is very strong even for the lowest gas volume fractions so that the  boundary layers do not react back on the bubbles.

On the other hand, the main distinction between the $\alpha=0.4\%$ and $\alpha=0.9\%$ cases and the cases with $\alpha \geq 1.4\%$  is that at low gas volume fraction the boundary layer on the side opposite to the bubble injection is not mixed by bubbles because almost no bubbles migrate to the other half of the setup, while at high $\alpha$ both boundary layers are mixed by bubbles.
Therefore the liquid (with almost no bubbles) flowing in the same direction as the cold boundary layer is more effective in transferring heat that the liquid flowing in the direction opposite to the movement of the hot boundary layer.
Finally, the observed differences in heat transport if the bubbles are injected close to the hot wall or close to the cold wall at $\alpha \geq 1.4\%$  are of the same order of magnitude as the variation of the Nusselt number for a constant gas volume fractions over the studied range of $Ra_H$, but are reproducible by repeating the measurements.
We also note that difference in the mixing mechanism close to the non-injection wall seems to be the cause of different exact scaling exponents of $\overline{Nu}$ with $\alpha$ observed for  $\alpha \leq 1.4\%$  for hot and cold wall injection.

\subsection{Local liquid velocity measurements} \label{velocity}

Liquid velocity measurements are performed by means of LDA with the goal of understanding the dynamics of the system and how it is affected by different bubble injection sides, heating, and different gas volume fractions, in order to relate it to our global heat transfer findings.

\begin{figure} 
\centering
\includegraphics[scale=0.9]{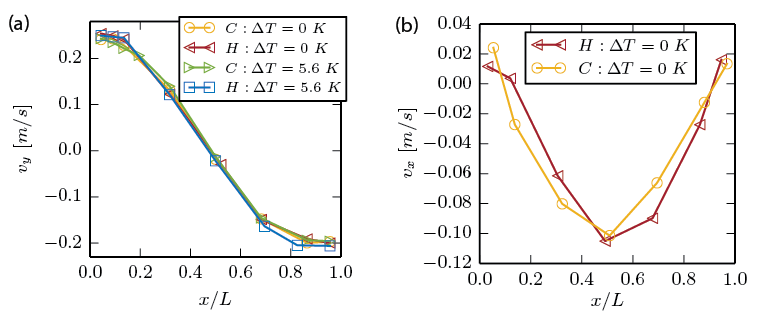}
\caption{(a) Vertical  velocity ($v_y$) profiles for a gas volume fraction  $\alpha =0.4 \%$, for different injection sides and different imposed wall temperature differences, (b) Horizontal  velocity ($v_x$) profiles for gas volume fraction  $\alpha =0.4 \%$, for different injection sides without the heating.}
\label{meanvel0800}
 \end{figure}

In the previous section we have seen that $\overline{Nu}$ is slightly higher for hot wall injection  when compared to cold wall injection for lower gas volume fractions. 
In order to examine this further we performed velocity measurements at half height of the setup in the direction normal to the heated and cooled walls for the case of lowest studied gas volume fraction $\alpha = 0.4\%$.
In figures \ref{meanvel0800}, we show the vertical and horizontal liquid velocity profiles for both hot-wall and cold-wall injections without the heating (cold side injection measurements are mirrored around $x/L=0.5$).    
The vertical velocity profile shows a high positive value at the injection side which is of the same order of magnitude as the bubble rising velocity, while the horizontal velocity is always negative indicating that the injected side is entraining the fluid from the opposite side. 
These findings  are in agreement with the results of  Roig \emph{et al.} (1998) \cite{roig1998experimental} who studied a turbulent bubbly mixing layer, which was produced by applying different inlet conditions of liquid velocity and gas volume fraction in two halves of a vertical square water channel. 
They found that even a very low difference between the gas volume fraction in the two halves of the setup induces strong acceleration of the fluid on the injection side of the mixing layer and a bending of the flow. 
Figure \ref{meanvel0800} (a) also shows the vertical velocity profiles for different injection sides with the heating ($\Delta T=\unit{5.6}{\kelvin}$). 
Results shown in figure \ref{meanvel0800} (a) indicate that heating and the side of injection do not have a significant influence on the mean velocity profiles. 
These findings do not comply with the ones for the global heat transfer possibly because the interaction of boundary layers with the co-current or counter-current bulk can only be captured by measuring velocity even closer to the heated and cooled wall, which is not possible due to strong reflections of the laser beam. 

\begin{figure*} 
\centering
\includegraphics[scale=0.9]{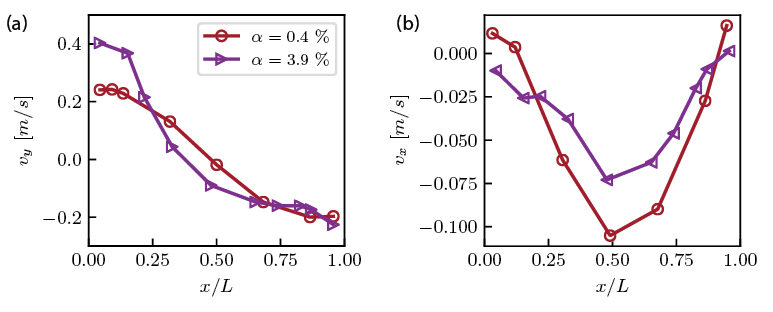}
\caption{Vertical velocity ($v_y$) and  horizontal  velocity ($v_x$) profiles for different gas volume fractions without the heating. The gas injection is performed at $x/L \leq 0.5$ .}
\label{meanvelboth}
 \end{figure*} 
Since $\overline{Nu}$ depends strongly on the gas volume fraction but not on $Ra_H$ we now look into the influence of gas volume fraction on the measured velocity profiles. 
In figure \ref{meanvelboth}, we show the vertical and horizontal velocity profiles at half height for gas volume fractions of $\alpha = 0.4\%$ and $\alpha = 3.9\%$. 
The horizontal velocity profiles are slightly higher in the centre of the setup for $\alpha = 0.4\%$.
This most likely occurs due to the presence of the bubbles on the side opposite to the injection side for  $\alpha = 3.9\%$, which leads to lower gradient of gas volume fraction at mid height for  $\alpha = 3.9\%$ and lower driving for horizontal motion.
Based on the profiles of the mean liquid velocity in the vertical direction a mean liquid rising velocity in the injection side of the setup $U$ can be estimated by averaging the velocity along the length of rising liquid layer.
This way we find that the mean velocity for $\alpha = 0.4\% $ is $\unit{0.16}{ \metre \per \second}$ and for $\alpha = 3.9\%$ is $\unit{0.25}{ \metre \per \second} $.
The mean velocity in the case  $\alpha = 0.4\% $ is comparable to the one observed in previous studies on mixing in inhomogeneous bubbly flows (where  $U = \mathcal{O} (10)\  \centi \metre$) \cite{almeras2016scalar,almeras2018mixing}, which further justifies the direct comparison between the global heat transport results and findings from these studies.

\section{Summary} \label{summary}
An experimental study on heat transport in inhomogeneous bubbly flow has been conducted. 
The experiments were performed in a rectangular bubble column heated from one side and cooled from the other, where the millimetric bubbles were injected through one half of the injection section, either close to the cold wall or close to the hot wall (see figure \ref{fig1}).
Two parameters were varied: the gas volume fraction (from $0.4\%$ to $5.1\%$) and the Rayleigh number (from $4 \times 10^9$ to $2.2\times 10^{10}$).

By characterising the global heat transfer we find that in the case of bubbles injected only through one half of the injection section, just as for homogeneous bubble injection, the Nusselt number is nearly independent on the Rayleigh number and increases with increasing gas volume fraction (see Figure \ref{Ra_Nu}).
However, the heat transfer enhancement is more prominent with inhomogeneously injected bubbles when compared to the same gas volume fraction and same range of $Ra_H$ of homogeneous injection, provided $\alpha<4\%$ (see Figure \ref{Nu_alpha}).
This finding can be explained by the multiple mixing mechanisms present in the setup, once a gradient of gas volume fraction is imposed. 
Namely, besides the bubble induced turbulence (BIT), the large-scale circulation of the liquid phase induced by inhomogeneous bubble injection leads to the occurrence of a shear layer between the fluid region injected with bubbles and its opposite side.
As previously observed by  Alm\'{e}ras \emph{et al.} (2016) \cite{almeras2016scalar} and Alm\'{e}ras \emph{et al.} (2018) \cite{almeras2018mixing} for $\alpha<3.5\%$, the different superimposed mixing mechanisms lead to enhancement of mixing, which  results in up to 1.5 times larger heat transport as compared to homogeneous bubble injection (see Figure \ref{Nu_alpha_HIdiv}).

Although the measurements of the velocity in the bulk show comparable profiles for hot wall and cold wall injections (see Figure \ref{meanvel0800}), the findings on global heat transfer indicate that the injection close to the hot wall induces stronger heat transfer enhancement for gas volume fraction lower than a critical value of $1.4\%$ (see Figures \ref{Ra_Nu} and \ref{Nu_alpha}).
At $\alpha \geq 1.4 \%$ we observe bubble-rich region near the non-injecting wall which promotes effective mixing near the thermal boundary layer at the wall. 
As a consequence the difference in the result on heat transport enhancement for the cases of hot wall and cold wall injection  at $\alpha \geq 1.4 \%$ is  diminished.
For  $\alpha < 1.4\%$ non-injection wall is not covered by bubbles, which means that in this range of  $\alpha$ the co-current flow of the liquid directed from the hot wall to the cold wall aids large-scale circulation and the heat transport enhancement.

For $\alpha>4\%$ the inhomogeneous injection causes lower heat transport enhancement than the homogeneous one.
We visually observe that with increasing gas volume fraction the instability of the bubble stream increases as well as the contribution of the shear-induced turbulence (SIT).
The velocity measurements show that the large-scale circulation gets stronger with increasing $\alpha$ as well (see Figure \ref{meanvelboth}).
Therefore the competition between BIT, SIT and the advection reduces the heat transport enhancement.

Lastly we comment on the generality of the obtained results on the heat transfer enhancement. 
When determining the generality of the results on the scaling of the Nusselt number with the gas volume fraction for lower studied gas volume fraction, one has to take into account the governing mixing mechanism in a inhomogeneous bubble column which depends on the aspect ratio (height over width) of the column and the size of the bubbles. 
If the bubble induced turbulence is the limiting mixing mechanism we expect the scaling of the Nusselt number to be the same, namely $Nu \propto \alpha^{0.45}$.
On the other hand, a significant change in the bubble size would affect the flow patterns as well.
Experimentally it is challenging to generate bubbles of significantly different sizes (in this study it is varied only from $\sim 1.8$ mm to $\sim 2.8$ mm), and it is difficult to predict how would the scaling be affected by it. 
We expect that the evolution of Nusselt number with the Rayleigh number is also general.
If the size of injected bubbles is comparable to the thickness of the thermal boundary layers (the thickness of the thermal boundary layer in the studied range of $Ra_H$ in the single phase case is estimated to be of order of few millimeters (see Gvozdi\'c \emph{et al.} (2018) \cite{gvozdic2018experimental}) and if the bubbles are injected close to the wall so that they perturb these boundary layers we expect that $Nu$ does not depend on $Ra_H$.
Unexplored effect of the changes in bubble sizes and the complexity of the interaction between different mixing mechanisms call for future investigations on this topic.

\section*{Acknowledgments}
This work is part of the Industrial Partnership Programme i36 Dense Bubbly Flows that is carried out under an agreement between Akzo Nobel Chemicals International B.V., DSM Innovation Center B.V., SABIC Global Technologies B.V., Shell Global Solutions B.V., TATA Steel Nederland Technology B.V. and the Netherlands Organisation for Scientific Research (NWO). 
Chao Sun acknowledges the financial support from Natural Science Foundation of China under Grant No. 11672156.
This work was also supported by The Netherlands Center for Multiscale Catalytic Energy Conversion (MCEC), an NWO Gravitation Programme funded by the Ministry of Education, Culture and Science of the government of The Netherlands.

\section*{References}

\bibliography{elsarticle-template.bib}
\linenumbers
\end{document}